# Electrical-controllable antiferromagnet-based tunnel junction

*Lei Han,[1,‡] Xuming Luo,[2,‡] Yingqian Xu,[2] Hua Bai,[1] Wenxuan Zhu,[1] Yuxiang Zhu,[1] Guoqiang Yu,[2] Cheng Song,[1,*] Feng Pan[1,*]*

[1]Key Laboratory of Advanced Materials (MOE), School of Materials Science and Engineering, Tsinghua University, Beijing 100084, China

[2]Beijing National Laboratory for Condensed Matter Physics, Institute of Physics, Chinese Academy of Sciences, Beijing 100190, China

Abstract: Electrical-controllable antiferromagnet tunnel junction is a key goal in spintronics, holding immense promise for ultra-dense and ultra-stable antiferromagnetic memory with high processing speed for modern information technology. Here, we have advanced towards this goal by achieving an electrical-controllable antiferromagnet-based tunnel junction of Pt/Co/Pt/Co/IrMn/MgO/Pt. The exchange coupling between antiferromagnetic IrMn and Co/Pt perpendicular magnetic multilayers results in the formation of interfacial exchange bias and exchange spring in IrMn. Encoding information states "0" and "1" is realized through the exchange spring in IrMn, which can be electrically written by spin-orbit torque switching with high cyclability and electrically read by antiferromagnetic tunneling anisotropic magnetoresistance. Combining spin-orbit torque switching of both exchange spring and exchange bias, 16 Boolean logic operation is successfully demonstrated. With both memory and logic functionalities integrated into our electrical-controllable antiferromagnetic-based tunnel junction, we chart the course toward high-performance antiferromagnetic logic-in-memory.

One of the key applications of spintronics is magnetic random-access memory (MRAM), which has attracted widespread research interest and found commercial use particularly in embedded systems, owing to its advantages of low power consumption, high endurance, and high integrability compared with other non-volatile memory technologies.[1-7] With logic operations integrated into MRAM arrays, in-memory computing that features great potential to fundamentally break through the Neumann bottleneck can be achieved, fitting well in futural big data process, artificial intelligence, Internet of Things, and edge computing.[8-12] The main operation principles of current MRAM can be concisely categorized into information storage, electrical reading, and electrical writing in magnetic tunnel junction devices, achieved through the orientation of magnetic moment, tunnel magnetoresistance (TMR),[13-15] and spin-transfer torque/spin-orbit torque (SOT),[16-19] respectively. Nevertheless, challenges such as the presence of ferromagnetic (FM) net moment, stray fields, and gigahertz intrinsic frequency make data in MRAM easily erased under magnetic disturbance, as well as limit FM-MARM towards higher integration density and faster operation speed.[20, 21] Antiferromagnetic (AFM) materials, characterized by their absence of net moment, zero stray field, and terahertz dynamics, naturally present prospects for solving these problems, gaining increasing attention as building blocks of high-performance AFM-MRAM for constructing logic-in-memory.[22-33] Similar to FM-MRAM, AFM-MRAM also requires achieving electrical write and electrical readout of the information state stored by the AFM moment within AFM tunnel junction.

Recently, the observation of TMR in AFM tunnel junction[34, 35] offers potential solutions to electrical readout, and the remaining critical challenge to achieve AFM-MRAM is to realize electrical manipulation of AFM moment in AFM tunnel junction. Previous research efforts largely focused on achieving electrical 90º, 120º or 180º switching of the Néel vector by SOT in antiferromagnets with special symmetry breaking or in antiferromagnet/heavy metal heterojunctions.[24, 36-39] Nevertheless, none of these phenomena was realized in AFM tunnel junction devices. An alternative and potentially more promising approach is harnessing the exchange interaction at the antiferromagnet/ferromagnet interface.[40-44] For instance, making use of the pinning effect of AFM IrMn on adjacent FM CoFeB (i.e., the exchange bias), the antiferromagnetic state of IrMn can be imprinted on the CoFeB layer. In this case, the SOT switching of IrMn and concomitant switching of exchange bias at the IrMn/CoFeB interface can be detected through traditional FM TMR in IrMn/CoFeB/MgO/CoFeB tunnel junction.[44] This work represents an important exploration of using AFM material as a functional layer in electrical-controllable tunnel junctions. However, this tunnel junction is essentially not an AFM tunnel junction, as there is no direct utilization of AFM electrode to generate tunneling resistance in the junction.

Our design paves the way of electrical-controllable AFM tunnel junction for in-memory computing by constructing an antiferromagnet-based Pt/Co/Pt/Co/IrMn/MgO/Pt tunnel junction. Information states are stored by the exchange-spring-induced AFM spin texture in IrMn, electrical reading is realized through AFM tunneling anisotropic magnetoresistance (TAMR),[23] and electrical writing is achieved by SOT switching of AFM exchange spring with high endurance. Moreover, when SOT switches exchange spring, the exchange bias between IrMn and Pt/Co perpendicular magnetic multilayers can be switched simultaneously, providing additional modulation dimension to enable 16 Boolean logic functions. The combination of information storage, reading, writing, and logic function in a single antiferromagnet-based tunnel junction device physically satisfies the requirement of AFM-MRAM as logic-in-memory.

The working mechanism of our electrical-controllable antiferromagnet-based tunnel junction Pt/Co/Pt/Co/IrMn/MgO/Pt is elucidated in detail in Figure 1. The bottom Pt layer functions as the spin source and the seed layer, and then the Co/Pt/Co trilayer with perpendicular magnetic anisotropy (PMA) is grown on the bottom Pt layer. IrMn serves as the AFM bottom electrode, facilitating the generation of TAMR across the MgO barrier. Strong interfacial exchange interaction will naturally form between AFM IrMn and PMA layers, giving rise to the two following key effects that underpin the fundamental physics of our design.

One effect is the tilting of certain rotatable IrMn moments (illustrated by red and blue arrows) from their initial orientations due to the exchange coupling with PMA moments (denoted by purple arrows), leading to the formation of exchange spring in IrMn. This exchange spring acts as a representation of information states and can be manipulated by reversing the magnetic field, resulting in high and low tunneling resistances (denoted as *H*-TAMR). These resistance states correspond to "1" and "0" information states, respectively (Figure 1a and Figure 1b). For electrical writing, when the switching current *I* is applied in the bottom Pt layer, SOT facilitates the switching of the adjacent PMA layer with the help of a small assistant field. Simultaneously, exchange spring that matches the downward orientation of PMA will be formed in IrMn, bringing about a low tunneling resistance representative of the "0" state (Figure 1c, denoted as *I*-TAMR).

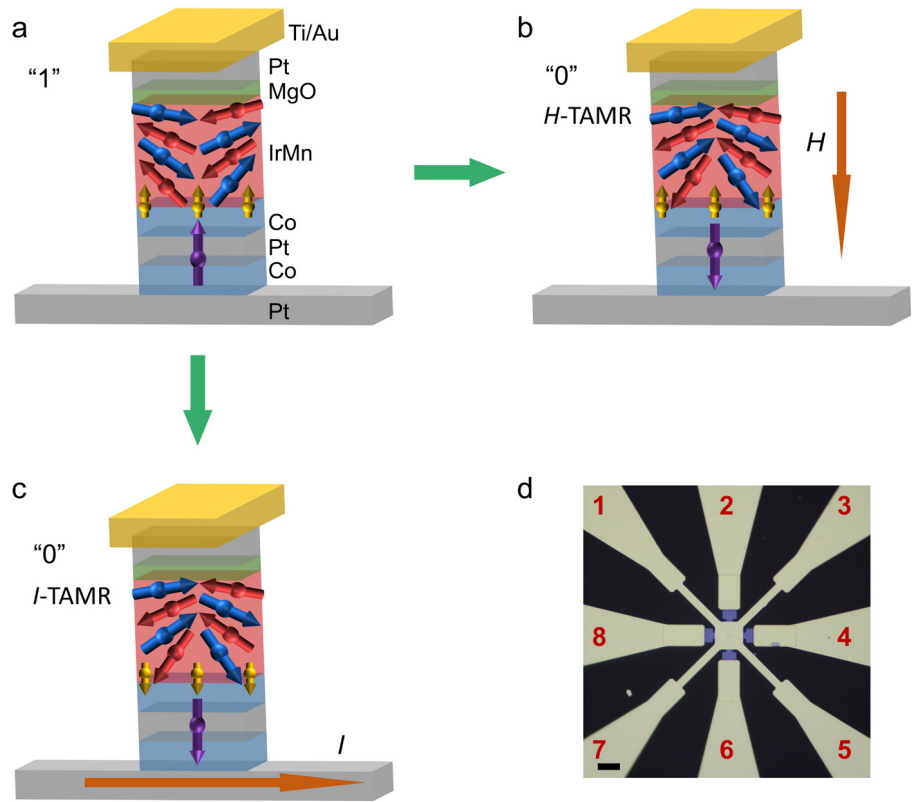

**Figure 1.** (a-c) Working principles and (d) optical microscopy image for our electrical-controllable antiferromagnet-based tunnel junction. The information states "1" and "0"

can be switched by either (b) magnetic field or (c) electrical current. The scalebar in (d) denotes 30 μm.

The other effect is that some non-rotatable IrMn moments (marked with yellow arrows) will interact with the adjacent PMA layer to bring about the exchange bias phenomena, reflected by a clear shift in the magnetic hysteresis or anomalous Hall hysteresis of PMA. Notably, non-rotatable IrMn moments can be simultaneously switched under SOT induced by current $I$[40, 42, 45, 46] (Figure 1c), but cannot be switched under magnetic field $H$ (Figure 1b). This unique property lays the foundation for the realization of logic functions.[45-47] Therefore, within a single antiferromagnet-based tunnel junction device, the exchange spring in IrMn enable information storage, electrical read, and electrical write, while the additional exchange bias effect facilitates the demonstration of logic function. Besides, because non-rotatable IrMn moments that determine the exchange bias are located at the interface between IrMn and PMA, the exchange bias switching will not generate TAMR across the MgO interface.

To achieve the aforementioned design, we first prepared Pt(12 nm)/Co(0.5 nm)/Pt(1 nm)/Co(0.5 nm)/IrMn(6 nm)/MgO(2.5 nm)/Pt(12 nm) multilayers on $Al_2O_3$(0001) substrates by magnetron sputtering at room temperature, ensuring high flatness (Supporting Information (SI), Experimental Section and Figure S1). Subsequently, this film was fabricated into AFM IrMn-based tunnel junction, where the junction area is a circle with a diameter of 5 μm. The optical microscopy image of this junction is shown in Figure 1d, featuring four top terminals of 1, 3, 5, 7, and four bottom terminals of 2, 4, 6, 8. The anticipated establishment of PMA in Co/Pt/Co, as well as the out-of-plane exchange bias between Co/Pt/Co and IrMn, was confirmed by the characteristic rectangular hysteresis loop of anomalous Hall effect (AHE) shifting towards the negative magnetic field direction (SI, Figure S2). For SOT switching of PMA, four bottom terminals are used to apply the switching current pulse and record the Hall voltage. For electrical-controlled TAMR measurements, a pair of bottom terminals serve as the writing channel, with TAMR obtained through the standard four-terminal method between the top and bottom terminals.

We start with SOT switching experiments of the PMA layer. It was performed by applying a series of writing current $I$ ranging from +36 mA to –36 mA and reversing back from –36 mA to +36 mA under an assistant field $H_x$ along the same channel of $I$ (SI, Experimental Section). Figure 2a shows a typical room-temperature SOT switching hysteresis of Hall resistance $R_H$ under $H_x$ of –0.2 kOe. The critical switching current $I_c$ for accomplishing full PMA switching is 35 mA. The switching ratio reaches 100% when compared with the AHE hysteresis (SI, Figure S2). When the polarity of the assistant field reverses, the SOT switching polarity also reverses (Figure 2b), consistent with basic symmetry requirements for SOT switching.[18, 19] Due to the interfacial exchange coupling, SOT switching of the bottom PMA layer naturally manipulates the exchange spring in IrMn simultaneously for the generation of AFM TAMR, as we discussed in the following.

We proceed to achieve electrical manipulation of this AFM IrMn-based tunnel junction. Similar series of writing current $I$ were applied along a pair of bottom terminals with an assistant field $H_x$, and TAMR was recorded simultaneously (SI,

Experimental Section). Figure 2c presents the result of the resistance area product (RA) under $H_x$ of –0.2 kOe, revealing a clear hysteresis. In-plane randomness of Néel vector does not quench the TAMR. It is because AFM TAMR arises from variations in interfacial density of states,[48, 49] which is originated from exchange-spring-induced changes of out-of-plane interfacial magnetic structure in IrMn. TAMR is calculated as (0.030 ± 0.006)%. This value can be increased to 0.36% by optimizing the IrMn thickness to be 5 nm (SI, Figure S3). For IrMn thicker than 12 nm, TAMR disappeared due to limited effective length of exchange spring. Further endeavors to enhance TAMR can be focused on realizing high-temperature growth of epitaxial IrMn(111) as bottom electrode or improving the crystal quality of MgO tunneling barrier. It is noteworthy that, due to the exchange spring is closely coupled with PMA moments, RA changes simultaneously with Hall resistance change during SOT switching of PMA (Figure 2a). Furthermore, the switching polarity of RA is opposite to that of SOT switching of PMA in Figure 2a, suggesting that this TAMR does not come from AHE generated by potential leakage current. Besides, temperature-dependence and bias-dependent RA measurements confirm the semiconductive characteristic of this junction, further proving the tunneling behavior of this junction without leakage (SI, Figure S4). Upon reversing $H_x$ to be 0.2 kOe, the polarity of RA hysteresis also reverses (Figure 2d), demonstrating that this RA change results from magnetic switching rather than resistive switching phenomena as possible artifacts. These results validate that the electrical manipulation of RA should be attributed to the electrical manipulation of the antiferromagnetic spin texture in IrMn, rooted in the SOT switching of exchange spring driven by the SOT switching of PMA.

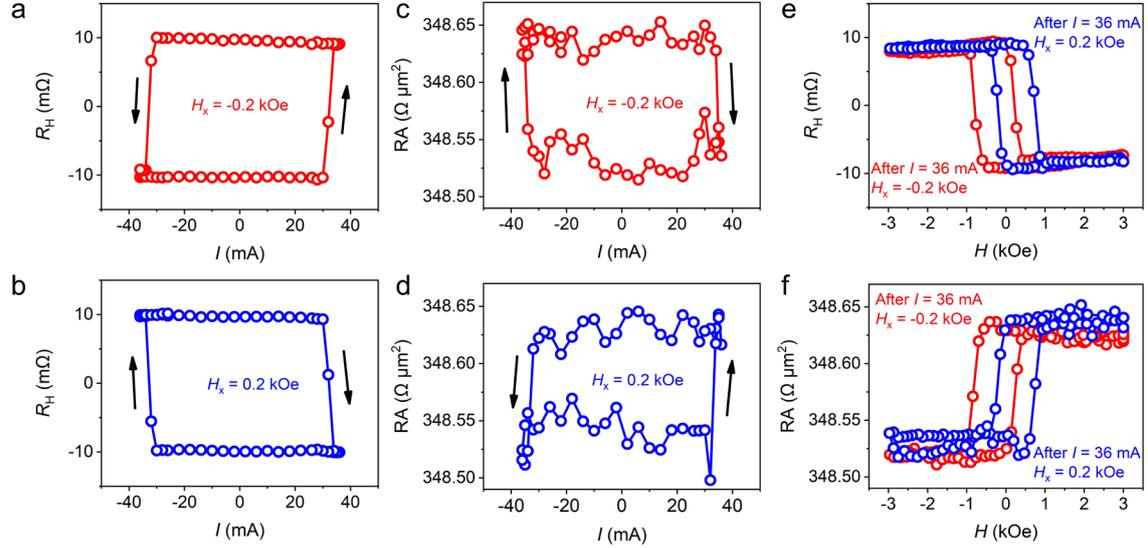

**Figure 2.** Electrical manipulation and electrical readout of the antiferromagnet-based tunnel junction. (a) SOT switching of the bottom PMA layer and readout out by the AHE under the assistance field $H_x$ of –0.2 kOe and (b) 0.2 kOe. (c) TAMR induced by the SOT switching of exchange spring under $H_x$ of –0.2 kOe and (d) 0.2 kOe. (e) AHE hysteresis and (f) $H$-dependent TAMR curve measured after SOT switching under $H_x$ of 0.2 and –0.2 kOe, indicating the switching of exchange bias.

Next, we move on to demonstrate the SOT switching of exchange bias. It is achieved by measuring the AHE hysteresis after SOT switching under sufficiently large $I$ with $H_x$ of ±0.2 kOe, as shown in Figure 2e. The AHE hysteresis shift towards opposite directions for $H_x$ of opposite polarities, indicating the switching of exchange bias under SOT. Here, the critical current for SOT switching of PMA and exchange bias are the same (SI, Figure S5), consistent with the previous results on similar Pt/Co/IrMn structure.[40] SOT determines the direction of PMA moments and simultaneously disturbs non-rotatable moments in IrMn for aligning them with PMA moments, manifesting as the switching of exchange bias.[40] Notably, there are also other works on SOT switching of exchange bias in IrMn/CoFeB/MgO[42] and Pt/IrMn/CoFeB[44] heterojunctions, where switching characteristics are different due to distinct spin source, stack structures, and interfacial exchange interaction.

SOT switching of exchange bias can also be reflected by measuring RA when sweeping out-of-plane magnetic field after SOT switching under sufficiently large $I$ with $H_x$ of ±0.2 kOe (Figure 2f). Firstly, the magnitude of RA change under magnetic field is equal to that under SOT. It matches the physical picture that both the magnetic field and SOT fully switch the PMA layer, which couples with AFM IrMn to manipulate its exchange spring, generating AFM TAMR. Secondly, the RA hysteresis also shifts towards reversing directions after SOT measurements under $H_x$ of opposite polarities, consistent with the shift of AHE hysteresis (Figure 2e) and revealing the switching of exchange bias. Thirdly, there is no vertical shift of RA hysteresis within the range of detection error, indicating that exchange bias switching does not contribute to TAMR. As a result, both exchange spring and exchange bias can be electrically switched by SOT in this AFM IrMn-based tunnel junction, opening avenues for achieving multifunctionalities such as spin logic in single AFM device.

The following will show the realization of 16 Boolean logic functions based on the abovementioned SOT switching of both exchange spring and exchange bias in a single AFM IrMn-based tunnel junction. The in-plane assistant field $H_x$, SOT switching current $I$, and out-of-plane switching magnetic field $H_z$ naturally serve as three inputs.

Additionally, an out-of-plane bias field $H_b$ is needed to describe the effect of exchange bias as the fourth input. Figure 3a shows the assignment of logic "0" and "1" for these four inputs, determined by their respective signs. The achievement of the logic function based on our AFM IrMn-based tunnel junction includes two steps. The first step is applying SOT switching current $I$ in a pair of bottom terminals under the assistant field $H_x$, resulting in a RA state as well as an exchange bias state as illustrated in Figure 2c, 2d, and 2f. The second step is simultaneously applying $H_z$ and $H_b$, which may or may not switch the RA state, depending on the exchange bias state. The final RA state after both steps will serve as the output signal, denoted as "0" or "1" for low or high RA, respectively.

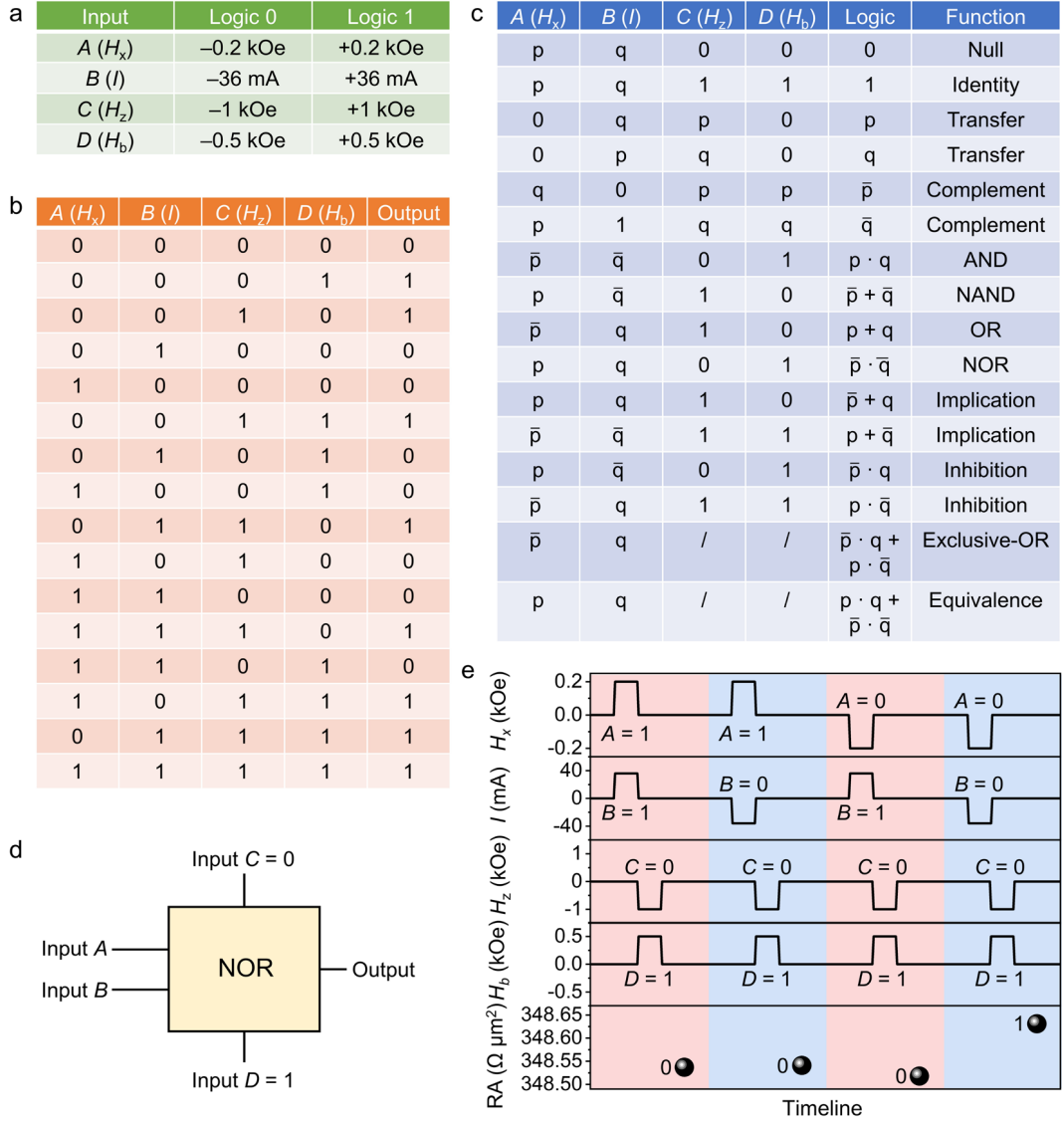

**Figure 3.** Realization of 16 Boolean logic functions in a single antiferromagnet-based tunnel junction. (a) Logic correspondence between 4 input variables. (b) All truth tables for the 4 input variables. (c) Assignment of the 4 variables to realize 16 Boolean logic functions. (d) A schematic for realizing the "NOR" function and (e) its detailed operation process with experimental results.

As shown in Figure 3b, we present a comprehensive list of all 16 possible combinations of the four inputs along with their corresponding output results, where the four inputs are denoted as $A$ ($H_x$), $B$ ($I$), $C$ ($H_z$), and $D$ ($H_b$) for simplicity. For instance, the first row of Figure 3b signifies that $H_x$ of –0.2 kOe, $I$ of –36 mA, $H_z$ of –1 kOe, and $H_b$ of –0.5 kOe result in a low RA state. Specifically speaking, in the first step, applying $I$ of –36 mA under $H_x$ of –0.2 kOe yields a high RA state (Figure 2c). In the second step, an out-of-plane magnetic field $H$ as $H_z$ plus $H_b$ of –1.5 kOe switches the exchange spring to give a low RA state (Figure 2f), corresponding to a "0" output.

It is evident that when both $H_z$ and $H_b$ are set to "0", a low RA state will always be obtained, regardless of the first step of SOT switching. It implies that by assigning $H_x$ and $H_b$ as input variables and fixing $H_z$ and $H_b$ to "0", the output will always be "0", representing a Null logic function. Similarly, we can design all 16 Boolean logic functions based on different configurations of the four inputs, as presented in Figure 3c. To make it clear, we discuss the relatively challenging Boolean logic function NOR as an example. In NOR function, an output of "1" is only attained if both inputs are "0". This function is achieved by fixing $C$ and $D$ to "0" and "1", respectively (i.e., fixing the out-of-plane magnetic field at –0.5 kOe) and using $A$ and $B$ as two input variables (Figure 3d). The physical meaning behind is that, the output of "1" can only be achieved by performing the first step under negative $H_x$ and negative $I$, in order to obtain a high RA state (Figure 2c) as well as a negative shift of RA hysteresis (Figure 2f). This NOR function is also verified experimentally, as depicted in Figure 3e.

Notably, for our AFM IrMn-based tunnel junction at present, magnetic field is involved as inputs for logic function. It would be more favorable for practical integrated circuits if all inputs are electrical signals. We propose a scheme to achieve this by introducing in-plane exchange bias in our tunnel junction, where this in-plane exchange bias can be redirected[43] by another current orthogonal to the current for PMA switching. Specifically, in-plane exchange bias ensures field-free SOT switching,[50] and the polarity of field-free switching is determined by the direction of in-plane exchange bias, providing additional modulation dimension for realizing 16 Boolean logic functions with all electrical inputs (SI, Figure S6).

Finally, we test the endurance of our AFM IrMn-based tunnel junction. We consecutively apply 1000 cycles of write current, alternating between +36 mA and –36 mA, under $H_x$ of 0.2 kOe, while simultaneously measuring RA (SI, Figure S7). As shown in Figure 4a and Figure 4b, the device exhibits no degradation after the completion of 1000 cycles. In fact, our tunnel junction can maintain the ability of electrical-controllable switching between high and low TAMR states even after $10^7$ cycles (SI, Figure S7). Besides, we continuously monitor RA for over 2000 seconds, and the high RA state remains stable. These findings prove the robustness of our AFM IrMn-based tunnel junction, demonstrating its capability for reliable electrical read and electrical write operations.

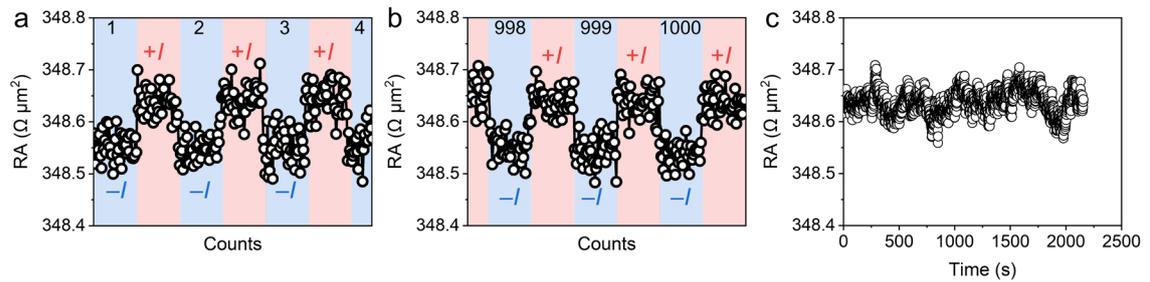

**Figure 4.** Device endurance tests. (a) The first 4 cycles and (b) The last 3 cycles of the electrical-controllable antiferromagnet-based tunnel junction during 1000 cycles test. (c) RA of the device for continuous read over 2000 seconds.

In summary, we design an electrical-controllable AFM IrMn-based tunnel junction, which successfully achieves both reliable memory functions and spin logic operations within a single device. Based on the exchange-spring-induced AFM spin texture in IrMn as the information storage medium, electrical writing is accomplished through SOT switching of exchange spring in AFM IrMn, while the information state is electrically readout by AFM TAMR signal. The simultaneous SOT switching of exchange spring in IrMn and exchange bias between IrMn and PMA layer enables the exploration of multifunctionality in this AFM IrMn-based tunnel junction, leading to the realization of 16 Boolean logic functions. Although the TAMR magnitude is limited, our design of AFM IrMn-based electrical-controllable tunnel junction with both memory and logic functions opens the possibility of developing electrical-controllable AFM logic-in-memory.

ASSOCIATED CONTENT

Supporting Information. The atomic force microscopy image, the anomalous Hall hysteresis, thickness dependence measurements, temperature-dependent and bias-dependent RA, exchange bias switching, design of 16 Boolean logic functions with all electrical inputs, and $10^7$ cycles of the tunnel junction (PDF).

AUTHOR INFORMATION

**Corresponding Author**

Cheng Song. Key Laboratory of Advanced Materials (MOE), School of Materials Science and Engineering, Tsinghua University, Beijing 100084, China. Email: songcheng@mail.tsinghua.edu.cn

Feng Pan. Key Laboratory of Advanced Materials (MOE), School of Materials Science and Engineering, Tsinghua University, Beijing 100084, China. Email: panf@mail.tsinghua.edu.cn

**Author Contributions**

L. H. X. L., and Y. X. prepared the samples and carried out transport measurements. This work was conceived, led, coordinated, and guided by C. S. and F. P. All the authors contributed to the writing of the manuscript. All authors have given approval to the final version of the manuscript. ‡These authors contributed equally.

**Notes**

The authors have no conflicts to disclose.

ACKNOWLEDGMENT

This work is supported by the National Key Research and Development Program of China (Grant No. 2021YFB3601301), National Natural Science Foundation of China (Grant No. T2394471, 52225106 and 12241404), and the Natural Science Foundation of Beijing, China (Grant No. JQ20010). Some devices were fabricated via an Ultraviolet Maskless Lithography machine (Model: UV Litho-ACA, TuoTuo Technology).